\documentclass[structabstract]{aa}  
\usepackage{graphicx}
\usepackage[round]{natbib}
\bibpunct{(}{)}{;}{a}{}{,} % to follow the A&A style
\usepackage{txfonts}
\def\cmmt{\rm {cm^{-2}}}

\def\s-1{\rm {s^{-1}}}

\begin{document}
   \title{Vibrationally excited HC$_3$N in NGC~4418}

%   \subtitle{}

   \author{F. Costagliola
          \inst{1}\fnmsep\thanks{F.C. wishes to thank the EU ESTRELA programme for support.}
          \and
          S. Aalto\inst{1}\fnmsep\thanks{S.A. wishes to thank the Swedish Research Council for grant support.}
          }

   \institute{Department of Radio and Space Science\\
			  Chalmers University of Technology\\
              Onsala Space Observatory\\
              SE-439 92 Onsala\\
              Sweden \\
              \email{francesco.costagliola@chalmers.se, saalto@chalmers.se}
             }

   \date{}

% \abstract{}{}{}{}{} 
% 5 {} token are mandatory
 
  \abstract
  % context heading (optional)
  % {} leave it empty if necessary  
   {}
  % aims heading (mandatory)
   {We investigate the molecular gas properties of the deeply
obscured luminous infrared galaxy NGC~4418. We address the excitation of the complex molecule HC$_3$N to determine whether its unusually luminous emission is related to the nature of the
buried nuclear source.}
  % methods heading (mandatory)
   {We use IRAM 30m and JCMT observations of rotational and vibrational lines of HC$_3$N to model the excitation of the molecule by means of rotational diagrams.}
  % results heading (mandatory)
   {We report the first confirmed extragalactic detection of vibrational lines of HC$_3$N.
We detect 6 different rotational transitions ranging from $J$=10--9 to $J$=30--29 in the ground vibrational state and obtain a tentative detection of the $J$=38--37 line. We also detect 7 rotational transitions of the vibrationally excited states $v_6$ and $v_7$, with angular momenta ranging from $J$=10--9 to 28--27. The energies of the upper states of the observed transitions range from 20 to 850 K. In the optically
thin regime, we find that the rotational transitions of the vibrational ground state can be fitted for two temperatures, 30 K and 260 K, while the vibrationally excited levels can be fitted for a rotational temperature of 90 K and a vibrational temperature of 500 K. In the inner 300 pc of NGC~4418, we estimate a high HC$_3$N abundance, of the order of 10$^{-7}$. }
 % conclusions heading (optional), leave it empty if necessary 
   {The excitation of the HC$_3$N molecule responds strongly to the intense radiation field and the presence of
 warm, dense gas and dust at the center of NGC~4418. The intense HC$_3$N line emission is a result of both high
abundances and excitation. The properties of the HC$_3$N emitting gas are similar to those found
for hot cores in Sgr~B2, which implies that the nucleus ($<$ 300 pc) of NGC~4418 is
reminiscent of a hot core. The potential presence of a compact, hot component (T=500 K) is also discussed.}

\titlerunning{HC$_3$N in NGC~4418}
   \maketitle
%
%________________________________________________________________

\section{Introduction}
Luminous infrared galaxies (LIRGs) are intriguing challenges to modern astronomy. They emit most of
their radiation in the infrared (IR) region of the spectrum in the form of  dust thermal continuum, and have typical
IR luminosities of L$_{IR}>$10$^{10}$ L$_\odot$. In many of these objects, the central power source responsible for the total energy
output is buried deep inside the dusty central region and has an origin that remains unclear. For instance, \citet{spoon07,nascent} suggest that dusty, compact LIRGs may represent the early obscured stages of either
active galactic nuclei (AGNs) or starbursts and thus play a fundamental role in galaxy formation and evolution.\\
Owing to the large column of intervening material, observations at IR or shorter wavelengths only detect
the surface of the optically thick nuclear regions, where molecular gas column densities can reach values that
exceed $N({\rm H}_2)=10^{24}$ $\cmmt$. Thus optical emission becomes completely absorbed by the
gas and dust, unless the source geometry allows the emission to escape. These high masses of gas and dust may also
cause an AGN to become Compton thick and hard X-rays to become absorbed by the intervening material, adding to problems in identifying the nuclear source.
At radio wavelengths, free-free emission from intervening ionized material may also obscure an 
AGN \citep[e.g., ][]{sanders_96}. 
This leads to great observational ambiguities that are reflected
by the classification of LIRGs varying considerably with the observed frequency band.
To produce the high central luminosity observed in many LIRGs requires either an AGN or a nuclear starburst (or a combination
of the two) to heat a large volume of dust. Dust temperatures in the inner hundred to a few hundred parsec vary
among galaxies and can range from 20-30 K to in excess of 170 K \citep[e.g., ][]{evans03,downes07}. Dissipation of turbulence, which
acts on the gas phase \citep[e.g., ][]{guesten93} may also indirectly contribute to the heating of dust.
Because of the aforementioned large amount of obscuring material,
the interplay between the possible energy sources remains difficult to ascertain by direct investigation.\\

Millimetre observations of molecular lines provide a potentially valuable tool for determining the effects of
the central power source on the interstellar medium (ISM) of LIRGs and indirectly constraining some of
their key properties, such as gas temperature, density, and chemistry. 
Chemical models \citep{meijerink07} show how X-ray dominated regions (XDRs), generally expected in the case of accretion onto a
central compact object, and photodissociation regions (PDRs), generated by large UV  fluxes from young stars, leave
different imprints in the ISM composition. The peculiar molecular chemistry of dense hot cores around nascent
stars was also described by \citet{viti08}. The gas temperature structure is also expected to be different
around an AGN  compared to a starburst. For example, in an XDR model bulk gas temperatures can be as high as 200 K, while in a starburst
the temperatures should be around 20-50 K \citep{meijerink07}.\\

Molecular lines surveys of LIRGs,  focused  mainly on high density tracers, such as HCN, HCO$^+$, HNC, and CS,
have been carried out by several groups \citep[e.g., ][]{krips08,gracia08,baan08,imanishi07,gao04,aalto02}.
However, even if these studies
provide unprecedented insights into the ISM of active galaxies, the interpretation of their results remains
debated and more sensitive tracers need to be found. \\In surveys of external galaxies (see Sect. \ref{sec:othergal}),  bright HC$_3$N\ $J$=10--9 line emission is found in
a subset of IR luminous galaxies.
The HC$_3$N\ line emission is a useful tracer of warm and dense regions and is extremely sensitive to a strong
IR-field because of its multiple, mid-IR bending modes. The molecule is often found to be abundant in Galactic hot cores \citep[e.g., ][]{devicente00} and can be easily destroyed by 
intense UV and particle radiation.
Thus HC$_3$N\  serves as a tracer of the gas (and dust) properties of galaxies with intense IR fields but where the dense gas is not too exposed to destructive radiation.

The edge-on, Sa-type galaxy NGC~4418 has one of the highest luminosities
in HC$_3$N (relative to HCN) found for an external
galaxy so far \citep{nascent,monje08}. Its unusual HC$_3$N emission was first reported by
\citet{nascent}. The inner region of NGC~4418 is deeply dust-enshrouded \citep{spoon_01} with mid-IR intensities
that are indicative of dust temperatures of 85 K \citep{evans03} inside a radius of 50 pc.
The IR luminosity-to-molecular gas mass ratio is high
for a non-ULIRG galaxy indicating that intense, compact activity is hidden behind the dust. 
Interferometric observations of HCN $J$=1--0 by \citet{imanishi04} show that the bulk of the dense
gas is contained in the inner 2$''$, corresponding to a region of 300 pc in diameter.
NGC~4418 is a FIR-excess galaxy with a logarithmic IR-to-radio continuum ratio ($q$) of 3
\citep{roussel03}. This excess may be caused by either a young pre-supernova starburst  or
 a buried AGN \citep{nascent,roussel03, imanishi04}.
The luminous HC$_3$N signature was interpreted as young starburst activity \citep{nascent}. However we need to perform more detailed observations and modeling of HC$_3$N emission of NGC~4418 to interpret its origin accurately .A tentative detection of a mm vibrational HC$_3$N\ line  \citep{nascent} led
us to explore the excitation and abundance of HC$_3$N\ in NGC~4418 and search for a model of
the dense gas properties in the inner 300 pc of the galaxy. Throughout this paper we assume that the
observed  HC$_3$N\ line emission emerges from a region smaller than or equal to the scale of the interferometric HCN
observations by \citet{imanishi04}. \\
In this paper, we report the first confirmed extragalactic detections
of vibrationally excited HC$_3$N in the LIRG NGC~4418 with the IRAM 30m telescope. We
report in total 13 different transitions
of HC$_3$N - including vibrationally excited levels - allowing us to compile a first model of the excitation and
abundance of HC$_3$N. We also report a tentative detection of the 345 GHz rotational $J$=38--37 line with the
JCMT telescope.
In Sect. 2, we present the observations and results in terms of line intensities. In Sect. 3,
we present the results in terms of HC$_3$N rotational diagrams and, in Sect. 4, we briefly discuss the 
interpretation of our results and future aims.

%__________________________________________________________________

\section{Observations}

The observations were carried out in December 2007 and July 2008 with the IRAM 30m telescope on Pico Veleta,
Spain. The coordinates of the observed position, taken from the NASA/IPAC  Extragalactic Database (NED), are  \textit{12h~26m~54.6s,~-00d~52m~39s (EQ 2000)} .  Calibration scans were taken every five minutes and pointing checked against nearby bright pointing sources every two hours. The nominal pointing accuracy was about 2$''$, much smaller than the typical full width at half maximum of our beam .
For both epochs, observations were performed in wobbler switching mode, with a throw of 120$''$ and 0.5 Hz phase, to optimize the baseline quality. The backends consisted of two filterbanks of 1~MHz and 4~MHz channel width, connected  to the 100 GHz and 230 GHz  receivers respectively.
The weather conditions in the 2007 run were excellent with precipitable water vapour of 1 mm or less.
 Even although the summer weather during the 2008 observations led to higher system temperatures, the poorer
 atmospheric transmission was counter-balanced by deeper integrations, leading to typical $rms$ noise
levels of 0.2 mK at 20 km/s resolution. \\
In the last observing run, the tuning of the IRAM 30m receivers was optimized to include the largest possible number of
vibrationally excited transitions, and the rotational transition of the vibrational ground state.
This allowed us to compare the new data with our previous observations of the vibrational ground state, to exclude possible calibration or baseline errors.
By comparing the two datasets, we conclude that the calibration errors are smaller than
or equal to 20\%.\\
Data analysis was performed with CLASS\footnote{http://iram.fr/IRAMFR/GILDAS/} and X-Spec\footnote{http://www.chalmers.se/rss/oso-en/observations/data-reduction-software} software packages. A first order baseline was subtracted from all spectra, which are shown in Fig. \ref{fig:spec}. The properties of
all the HC$_3$N observed transitions are reported in Table \ref{tab:lines}, together with main beam efficiencies, transitions
probabilities and energies of the upper states.

\section{Results}

\subsection{Detected lines} 
\label{sec:detected}
The rest-frame frequencies of the observed transitions were taken from the NIST database of {\it Recommended Rest Frequencies for
Observed Interstellar Molecular Microwave Transitions}\footnote{http://physics.nist.gov/PhysRefData/Micro/Html/contents.html} and cross-checked with laboratory and theoretical values
from \citet{yamada86}. The notation for the vibrational quantum numbers is in the form
{$J_u$-$J_l$ $(v_6,v_7)$}, following \citet{yamada86}.

The interaction between the bending angular momentum of the vibrationally excited states and the rotational angular momentum of
the molecule, leads to a $l$-splitting of the levels. Each vibrational state is thus split into two levels, labelled
 $e$ or $f$ depending on the wavefunction's parity properties. These parity labels are also shown in Table \ref{tab:lines}.  
A useful reference for labelling of doubled levels in linear molecules is \citet{brown75}.

We detected the HC$_3$N rotational transitions $J$=10--9, 16--15, 17--16, 25--24, 28--27, and 30--29. For the
$J$=10--9, 17--16, 25--24, and 28--27 lines, we also detected rotational transitions of the $v_7$=1 vibrationally excited levels.
For the $v_6$=1 lines, we generally have upper limits, the only clear detection being in the $J$=25--24 band. 
The observed HC$_3$N transitions were selected to minimize blends by major species. We excluded the possibility of blending by radio recombination lines as well as by other molecules (e.g., methanol). The lack of methanol emission is interesting and will be discussed in an upcoming paper.

\begin{figure*}
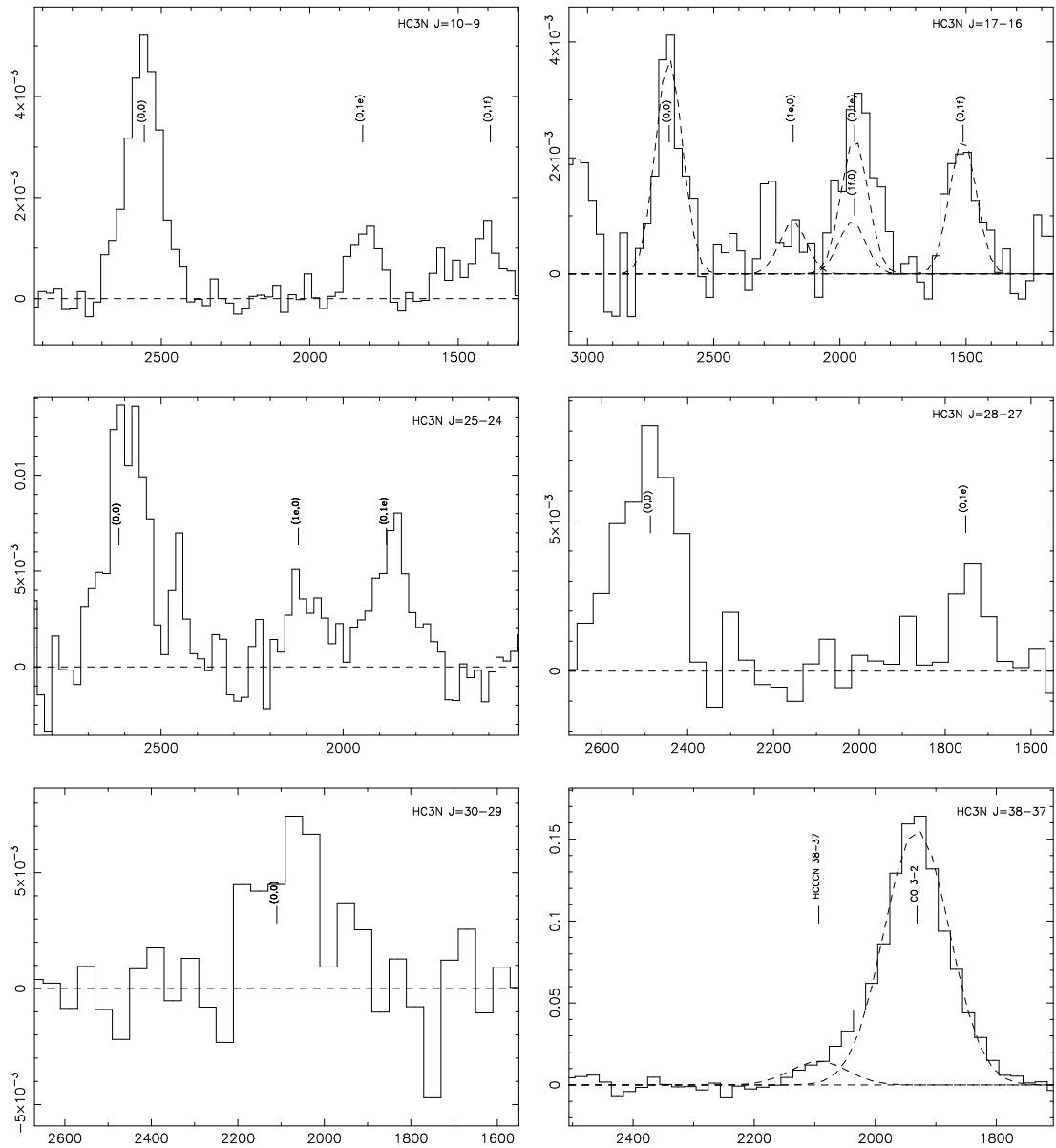

   \centering
   \includegraphics[height=.40\textwidth, angle=-90]{hc3n10-9.ps}
   \includegraphics[height=.40\textwidth, angle=-90]{hc3n17-16.ps}\\
   \includegraphics[height=.40\textwidth, angle=-90]{hc3n25-24.ps}
   \includegraphics[height=.40\textwidth, angle=-90]{hc3n28-27.ps}\\
   \includegraphics[height=.40\textwidth, angle=-90]{hc3n30-29.ps}
   \includegraphics[height=.40\textwidth, angle=-90]{hc3n38-37.ps}
   \caption{Spectra of HC$_3$N in NGC~4418. On the X axis, the velocity scale is shown in km/s. 
The intensity is T$_A^\star$ in Kelvin, not corrected for main beam efficiency and beam size. The central tuning frequencies for each observed band are listed in Table \ref{tab:lines}.  In the two cases {\it(top right, bottom right)} where significant blending is present, we show the Gaussian fits as dashed lines.  See text in Sect. \ref{sec:detected} for discussion about line identification. }
              \label{fig:spec}%
    \end{figure*}

Thanks to the relatively narrow line widths of about 120 km/s, line confusion is not a problem in most of the spectra. The most
crowded situation is found in the $J$=17--16 band, where multiple Gaussian fits were used to separate the $v_7$=1e and $v_6$=1f lines.
Since {\it e} and  {\it f} transitions come from the {\it l}-doubling of the same bending mode, we can assume that they have the
same intensity. Fixing the $v_7$=1$e$ intensity to the more accurately determined $v_7$=1$f$ value (see Fig. \ref{fig:spec}), we were able
to separate the $v_6$=1$f$ contribution.  We note that the fitted value for $v_6$=1$f$ is consistent with the measured upper limit
for the $v_6$=1$e$ line.  

The molecule appears to be highly excited, with very bright emission even for large upper angular momenta and energies. After realizing this, we checked spectra of other molecules at higher frequencies to search for HC$_3$N contamination finding an
asymmetry in the CO 3--2 spectrum at 345 GHz, taken by Monje and Aalto in 2007 with the James Clark Maxwell Telescope in Hawaii
(see Fig \ref{fig:spec}, bottom right). We interpreted this asymmetry as the HC$_3$N $J$=38--37 line, whose fitted values
are also reported in Table \ref{tab:lines}. This tentative detection was not included in our analysis.

%  Two column figure 
%  Observed Spectra

%
%
%______________________________________________________________
\begin{table*}
\caption{Observed HC$_3$N Lines Properties}  
\label{tab:lines}      
\centering         
\begin{tabular}{r c c c c c c c c c c}     % 7 columns 
\hline\hline       
Transition & Rest Frequency & Tuning Frequency & HPBW$^\mathrm{a}$ & $\int{T_A^\star dv}$ & T$_{b}^\mathrm{b}$ & $\Delta V^\mathrm{c}$ & $\eta_{mb}^\mathrm{d}$ & $A
_{u\ell}^\mathrm{e}$ & E$_u^\mathrm{f}$ & Ref.$^\mathrm{g}$ \\
J$_u$-J$_l$ $(v_6,v_7)$ & [GHz] & [GHz] & [$"$] & [mK km/s] & [K] & [km/s] & & [s$^{-1}$] & [K] & \\  
\hline     
& & & & & & & & &\\               
10--9 (0,0) & 90.979 & 91.114 & 27 & 711 $\pm$ 20 & 1.5 & 134 & 0.77 & 5.8e-05 & 24.1 & IRAM 07-08 \\
     (1e,0) & 91.128 & $''$ & 27	& $<$70 & - & - & 0.77 & 5.8e-05 & 736.6	& IRAM 07-08 \\
     (0,1e) & 91.202 & $''$ & 27 & 159 $\pm$  17 & 0.4  & 108  & 0.77 & 5.8e-05 & 342.6	& IRAM 07-08 \\
     (0,1f) & 91.333 & $''$ & 27 & 155  $\pm$ 14 & 0.4  & 110 & 0.77 & 5.8e-05 & 342.6 & IRAM 07-08 \\
16--15 (0,0) & 145.561 & 145.561 & 17 & 1700 $\pm$ 80 & 1.7  &	130 & 0.70 & 2.4e-04 & 59.4 & Aalto 2007 \\
17--16 (0,0) & 154.657 & 154.948 & 16 & 469 $\pm$ 52 & 0.5  & 117 & 0.67 & 2.9e-04 & 67.3 & IRAM 07-08 \\
      (1e,0) & 155.032 &$''$  & 16 & $<$150 & -  & - & 0.67 & 2.9e-04 & 779.2 & IRAM 07-08 \\
      (0,1e) & 155.037 &$''$  & 16 & 316 $\pm$  59 & 0.3  & 141 & 0.67 & 2.9e-04 & 385.2 & IRAM 07-08 \\
      (0,1f) & 155.259 &$''$  & 16 & 316 $\pm$  59 & 0.3  & 141 & 0.67 & 2.9e-04 & 385.3 & IRAM 07-08 \\
      (1e,0) & 154.911 &$''$  & 16 & $<$150 & -  & - & 0.67 & 2.9e-04 & 779.2 & IRAM 07-08 \\
25--24 (0,0) & 227.418 & 227.757 & 11 & 2020 $\pm$  140 & 1.1  & 140 & 0.54 & 9.3e-04 & 143.0 & IRAM 12-07 \\		
      (1e,0) & 227.793 & $''$ & 11 & 480 $\pm$ 110 & 0.3  & 125 & 0.54 & 9.3e-04 & 853.9 & IRAM 12-07 \\	
      (0,1e) & 227.977 & $''$ & 11 & 928  $\pm$ 120 & 0.6 & 125 & 0.54 & 9.3e-04 & 459.9 & IRAM 12-07 \\
28-27 (0,0) & 254.699 & 254.699 & 10 & 1720 $\pm$  100 & 0.9  & 141 & 0.47	& 1.3e-03 & 178.6 & IRAM 12-07 \\
28-27 (0,0) & 254.699 & 255.018 & 10 & 1315 $\pm$  100 & 0.7 & 159 & 0.47 & 1.3e-03 & 178.6 & IRAM 07-08 \\
	 (1e,0) & 255.119 &$''$  & 9.7 & $<$200 & -  & - & 0.47 & 1.3e-03 & 889.1 & IRAM 07-08 \\
      (0,1e) & 255.324 & $''$ & 9.7 & 327 $\pm$ 82 & 0.3  & 83 & 0.47 & 1.3e-03 & 495.2 & IRAM 07-08 \\
30-29 (0,0) & 272.884 & 272.884 & 9	& 1300 $\pm$  200 & 0.5  & 182	& 0.44 & 1.6e-03 & 204.6 & IRAM 07-08 \\
38--37 (0,0) & 345.609 & 345.609 & 14 & 1930 $\pm$  250$^{\mathrm{h}}$ & 1.8  & 120 & 0.63 & 3.3e-03 & 326.0 & JCMT 2007 \\ 	

& & & & & & & & \\
\hline
\end{tabular}
\begin{list}{}{}
\item[$^{\mathrm{a}}$] Half power beam width.
\item[$^{\mathrm{b}}$] Brightness temperature, assuming a 2$''$ source size.
\item[$^{\mathrm{c}}$] Full width at half maximum from Gaussian fit.
\item[$^{\mathrm{d}}$] Main beam efficiency.
\item[$^{\mathrm{e}}$] Einstein's transition coefficient.
\item[$^{\mathrm{f}}$] Energy above ground of the upper level.
\item[$^{\mathrm{g}}$] Telescope and epoch. 
\item[$^{\mathrm{h}}$] From Gaussian fit to the CO 3-2 spectrum, see Fig \ref{fig:spec}.
\end{list}
\end{table*}

\subsection{Line widths}

The observed line widths vary from 80 km/s for the $J$=28-27, $v_7=1$ transition to 182 km/s for $J$=30-29. If we exclude these two extreme cases, the remaining lines show comparable line widths, with a mean value of 125 km/s and a standard deviation of 12 km/s that is smaller than the spectral resolution, which for our data is not less than 20 km/s. Line widths do not show significant trends when compared to the energy of the upper level of observed transitions and thus suggest that the emission is originating in the same dynamical region inside the galaxy. 
The observed line widths are comparable to those inferred from interferometric observations of other dense gas tracers, such as HCO$^+$ and HCN \citep{imanishi04}. Unfortunately, these observations can barely resolve the 2$''$ emitting region and do not offer much information about the spatial distribution and dynamics of the dense gas.\\ 
It is remarkable that an edge-on galaxy exhibits these narrow emission lines. This might imply that the dense gas is mostly  outside the very nuclear region. High resolution observations would help us to determine the velocity field of the central gas.

\subsection{Population diagram}

The observed line intensities were combined in a population diagram (see Fig. \ref{fig:rot}) to attempt a first LTE
analysis of the emission. An extensive description of the population diagram method for deriving interstellar gas properties is
given by \citet{popdiag}. 
From the basic theory of molecular emission, it can be shown that the population of a certain energy
level is given by 
 \begin{equation}
%\begin{center}
\label{eq:pop}
N_u=\frac{8\pi k \nu^2 W}{h c^3 A_{u\ell}}\left(\frac{\tau}{1-e^{-\tau}} \right),
%\end{center}
\end{equation}
where W is the integrated brightness temperature. This is related to the observed integrated T$_A^\star$ listed in Table \ref{tab:lines}
by

\begin{equation}
%\begin{center}
W=\int{\frac{T_A^\star}{\eta_{mb}}\frac{\theta_s^2+\theta_b^2}{\theta_s^2}dv}=
\frac{\theta_s^2+\theta_b^2}{\eta_{mb}\theta_s^2}\int{T_A^\star dv},
%\end{center}
\end{equation}
where $\theta_s$ and $\theta_b$ are the angular sizes of the source and the telescope beam respectively. In our calculations we assume $\theta_s$=2$''$,
which is the upper limit to the  size of the HCN 1-0 emitting region found by \citet{imanishi04} with interferometric observations.  We note that we consider source-averaged quantities, since the size of the source is much smaller than the beam and a beam-averaged approach would lead to an underestimate of the true brightness temperature and column density.
Einstein's transition probability coefficients $A_{u\ell}$ were taken from the {\it Cologne Database of Molecular Spectroscopy}\footnote{http://www.astro.uni-koeln.de/cdms/} (CDMS).\\
At LTE, we have 
\begin{equation}
\label{eq:lte}
N_u=\frac{N}{Z}g_ue^{-E_u/T},
\end{equation}
where $N$ is the total number density of molecules and $Z$ is the partition function. For $kT>hB_0$, i.e.,  for warm molecular gas, we can approximate
$Z\simeq kT/hB_0$, the rotational constant $B_0$ being 4.55~GHz for HC$_3$N. \\
Plotting the observed values of equation (\ref{eq:pop}) versus the energy of the upper level $E_u$, we can find the excitation temperature
of the molecule by fitting a line to the natural logarithm of Eq. (\ref{eq:lte}). The excitation temperature is given by
the slope of the fitted line.\\

The population diagram for the observed transitions is shown in Fig. \ref{fig:rot}. The line intensities were fitted using a minimum-$\chi^2$ method. The free parameters of the fit were column density and excitation temperature, which could vary, respectively, within the ranges 10$^{12}$-10$^{16}$ cm$^{-2}$ and 10-600 K. The fit includes opacity corrections and results in an optically thin scenario, with typical optical depths of 0.04 for $v$=0 and 0.2 for the $v_7$=1 transitions. A discussion about opacity effects for larger molecular column densities can be found in Appendix~\ref{sec:opacity}.\\The excitation of the molecule is described by four temperature components.

\subsubsection{The rotational temperatures}

The $v$=0 levels cannot be fit by a single line and
clearly exhibit two temperature components at 29 K (for E$_u<$100 K) and 265 K (for larger E$_u$).
The $v_7$=1 transitions have an excitation temperature of 91 K.
These are all rotational temperatures referring to the excitation of angular momentum states of the molecule. 

\subsubsection{The vibrational temperature}
When we compare the populations of different vibrational levels with the same $J$, we determine the vibrational
temperature, which describes the excitation of the vibrational modes. This can be achieved for the $J$=25--24 band,
for which we have both $v_6$=1 and $v_7$=1 lines. The resulting vibrational temperature is 519 K.\\
The accuracy of the excitation temperatures and column densities estimated by population diagram fitting were tested by Monte Carlo simulations. These results assign a confidence limit of about 30\% and 20\% to the derived temperatures and column densities, respectively.

\begin{table}[!h]
\caption{Excitation properties of HC$_3$N bending modes from \citet{wyr99}. Notice the large critical densities, see discussion
in the text.}             % title of Table
\label{tab:vib}      % is used to refer this table in the text
\centering                          % used for centering table
\begin{tabular}{c c c c c}        % centered columns (4 columns)
\hline\hline                 % inserts double horizontal lines
Transition & $\lambda_{ex}$ & A$_{u\ell}$ & n$_{cr}$ & vibration \\    % table heading 
& [$\mu$m] & [s$^{-1}$] & [cm$^{-3}$] &  \\
\hline                        % inserts single horizontal line
& & & &\\
$v_6$=1-0 & 20 & 0.15 &  3$\times$10$^{11}$ & CCN bend\\
$v_7$=1-0 & 45 & 6$\times$10$^{-4}$ & 4$\times$ 10$^{8}$ & CCC bend\\
& & & &\\
\hline                                   %inserts single line
\end{tabular}
\end{table}

\begin{figure}[h]
\centering
\includegraphics[width=.5\textwidth]{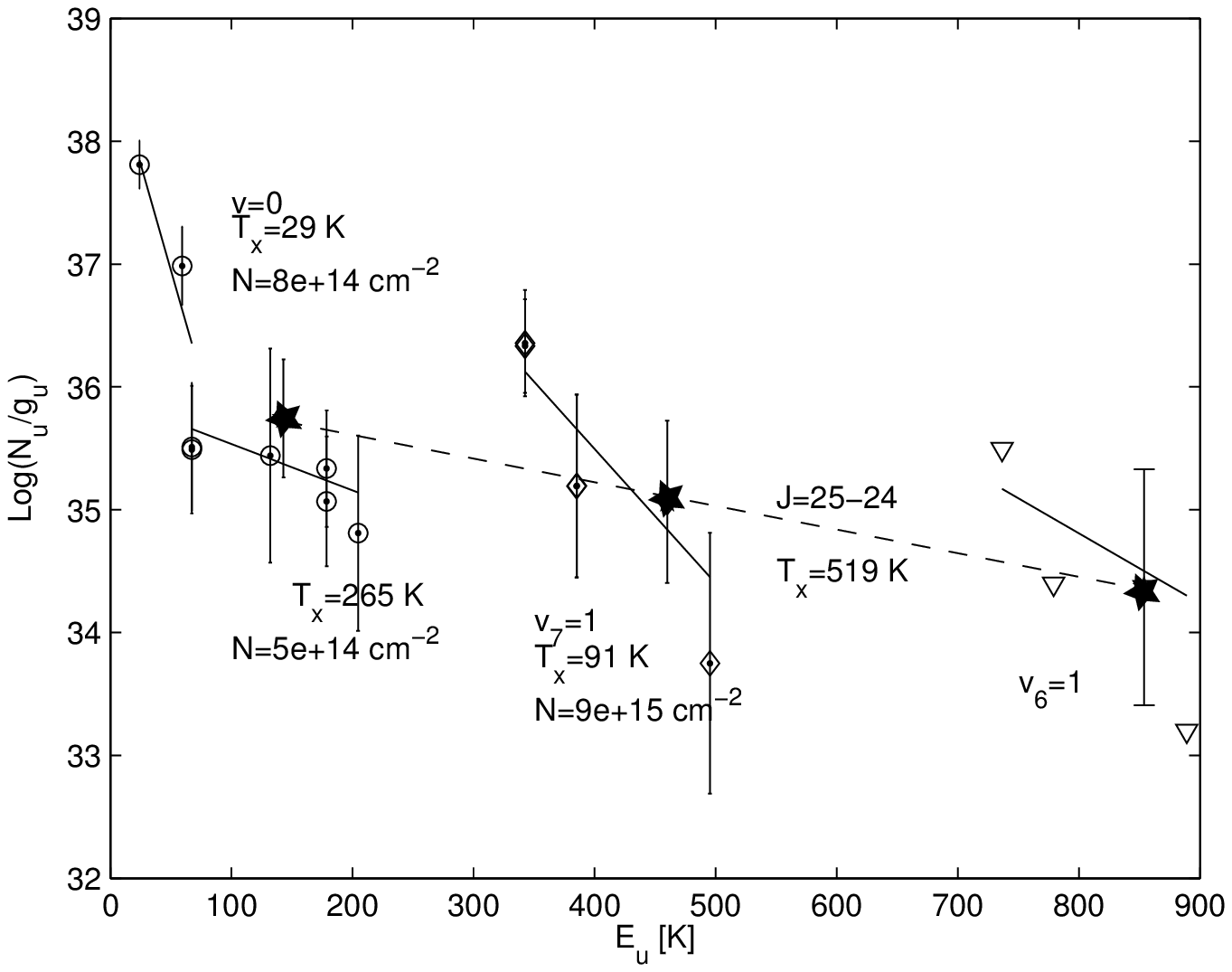}
\caption{Population diagram for all the observed HC$_3$N lines in NGC~4418. The 3-$\sigma$ error bars include uncertainties in the line intensity and beam size. The solid lines represent the fit of the level population for $v$=0 ($circles$), $v_6$=1 ($triangles$), and $v_7$=1 ($diamonds$). The dashed line fits the $v$=0, $v_6$=1, and  $v_7$=1 vibrational states of the $J$=25 level (marked with a $star$). For the $v_6$=1 lines, we note that the only detection is for $J$=25--24, the others being only upper limits. The vibrational ground state was fitted with two temperature components. The fit includes opacity corrections, as described in Appendix~\ref{sec:opacity}. Maximum opacities are 0.04 for $v$=0 and 0.2 for $v_7$=1.}
\label{fig:rot}
\end{figure}

\section{Discussion}

In NGC~4418, the emission of HC$_3$N is the strongest ever detected in an extragalactic source
with a HC$_3$N (10--9)/ HCN (1--0) line ratio of about 0.4. The excitation of HC$_3$N is also unusual with a large number of
vibrationally excited lines detected, suggesting that the excitation of the molecule is strongly affected by radiation.

\subsection{Models of the excitation of HC$_3$N}

If we assume that HC$_3$N\ is optically thin, multiple temperature 
components must contribute to the emission. This is possible in a galaxy characterized by
steep temperature gradients in its inner regions. We first examine how the radiation field
may affect the excitation of the HC$_3$N molecule, and then discuss the possible origin
of the temperature components.

\subsubsection{IR pumping of HC$_3$N}
\label{sec:pump}

\begin{figure}[b]
\centering
\includegraphics[width=.4\textwidth]{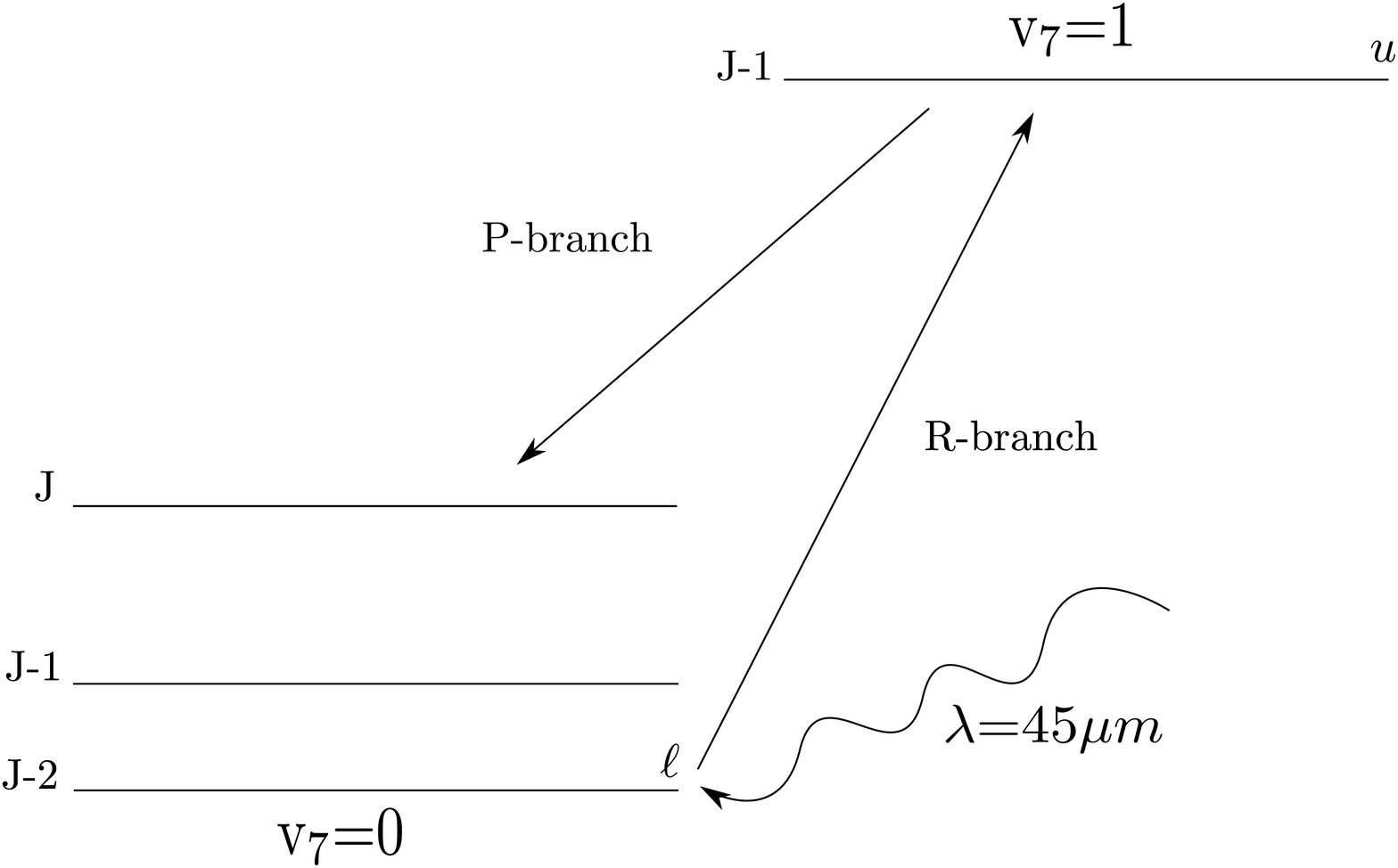}
\caption{Radiative pumping of HC$_3$N rotational levels via vibrational IR transitions.}
\label{fig:pump}
\end{figure}

The ground vibrational state of HC$_3$N can be excited by IR radiation at $\lambda$=45 $\mu$m or $\lambda$=20 $\mu$m, respectively
to a $v_7$=1 or $v_6$=1 state (see Fig. \ref{fig:pump}). The excited state can then decay through the R-branch ($\Delta J$=-1) or
P-branch ($\Delta J$=+1) to populate a $v$=0 level, with a resulting selection rule $\Delta J$=2 \citep{goldsmith81}. The level population can
be estimated from Boltzmann's equation 

\begin{equation}
\frac{n_u}{n_\ell}=\frac{B_{\ell u}I}{B_{u\ell}I+A_{u\ell}}=\frac{g_u}{g_\ell}e^{ -\frac{h\nu}{k T}},
\end{equation}
where $\ell$  and $u$ refer respectively to the ground and excited vibrational state.
Assuming $g_\ell B_{\ell u}=g_u B_{u \ell}$, the maximum pumping rate is given by
\begin{equation}
B_{\ell u}I=\frac{A_{u \ell}}{e^{\frac{h \nu}{k T}}-1}\equiv P_{\ell u}.
\end{equation}
This has to be compared to the collisional excitation rate $C_{J-1\rightarrow J}=n(H_2)q_{J-1\rightarrow J}$ for the rotational levels of the ground vibrational state. The previous formula gives an upper limit for the {\it pumping critical density} of $n_{cp}=P_{\ell u}/q_{J-1\rightarrow J}$. \\For $n(H_2)<n_{cp}$ the radiative pumping process is more efficient than collisions in exciting the molecule.

The collisional coefficients $q_{J-1\rightarrow J}$ of the first 20 rotational levels of HC$_3$N, for different gas temperatures, can be found in
the Leiden Atomic and Molecular Database. We considered an impinging IR flux coming from an optically thick dust source at 85 K.
We used this temperature for determining both pumping rates and  collisional coefficients. Typical pumping rates are of the order
of 10$^{-5}$ s$^{-1}$ for both $\lambda$=20 $\mu$m and $\lambda$=45 $\mu$m across all the range of upper state angular momenta.
The collisional coefficients at 85 K do not show significant variations for different transitions and have a value of about
10$^{-10}$ cm$^{-3}$ s$^{-1}$. The resulting pumping critical density is $n_{cp}\simeq 10^5$ cm$^{-3}$. 
The density structure of the HC$_3$N emitting gas in NGC~4418 remains unknown
and therefore we cannot provide a quantitative estimate of the impact of pumping
on the ground state rotational levels. However, for a clumpy medium where dense
clouds are surrounded by low-density diffuse molecular gas, the pumping would act directly
on the diffuse medium raising the global intensity and rotational temperature of
the emission.

\subsubsection{The cool, 30 K, and warm 90 K components}

The 90 K rotational temperature of the v$_7$ bending ladder may originate in dense gas within the inner
$0.5''$. The dust temperature of 85 K found by \citet{evans03} agrees well with the rotational temperature,
which may be because either  the IR radiation field dominates the excitation, or the
gas and dust are in thermal equilibrium and the rotational temperature reflects the gas kinetic temperature. 

The low temperature, 30 K, component could be coming from a more extended phase, at a greater distance from the 
warm dust in the center. The existence of this cool component is inferred from the lower transition v=0 lines
and can be confirmed by observations of low-$J$ lines with, for example, the Effelsberg telescope.

\subsubsection{The hot, 260 K and 500 K, components}

As we can see in Fig. \ref{fig:rot}, the estimated rotational temperature of the high-$J$ transitions is higher than the 30 K found for the low-$J$s, by almost one order of magnitude. This change in the slope of the rotational diagram cannot be explained by opacity effects and must be investigated further. Concave population diagrams are commonly found in Galactic molecular observations and are often attributed to multiple temperature layers. Temperatures of 300 K are typical of Galactic hot cores around sites of massive star formation (see Sect. \ref{sec:inthegalaxy}), the excitation temperature found could thus be reflecting the true gas kinetic temperature. This is in agreement the detections of the 14 $\mu$m absorption of HCN, for which \citet{lahuis07} find gas temperatures of about 300 K in the core of NGC~4418.\\
The vibrational temperature of the $J$=25--24 lines is about 500 K. The bending modes v$_6$ and v$_7$ both have critical densities greater than 10$^8$ cm$^{-3}$ (see Table \ref{tab:vib}) and are thus probably radiatively excited. The derived vibrational temperature then reflects the temperature of a radiation field.

\subsubsection{Radiative excitation of high-J levels}

The high temperature of the high-$J$ rotations may also be caused by radiative excitation, that becomes more and more efficient as the angular momentum of the upper state increases. 
The critical density of a $J \rightarrow J-1$ transition can be written as 
\begin{equation}
n_c=\frac{A_{J,J-1}}{q_{J,J-1}}=\frac{64\pi^4}{3hc^3}\nu^3\mu^2\frac{J+1}{2J+3}\frac{1}{q_{J,J-1}},
\end{equation}
where $\nu$ is the line frequency and $\mu$ is the electric dipole moment of the molecule.
For $n(H_2)<n_c$, the excitation temperature is determined by the radiation field and not the gas kinetic temperature.
\begin{figure}
\centering
\includegraphics[width=.4\textwidth]{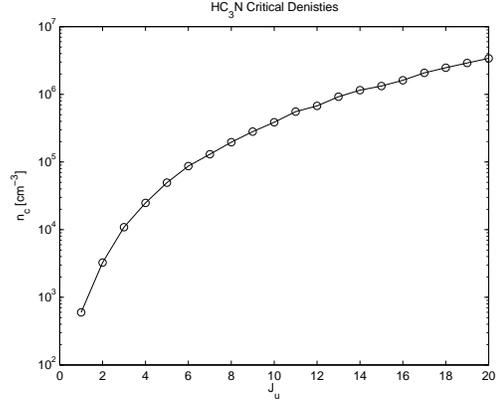}
\caption{Critical densities for the first 20 rotational levels of HC$_3$N. Emission and collisional coefficients from Leiden Atomic and Molecular Database.}
\label{fig:nc}
\end{figure}     
The critical densities of the first 20 rotational transitions of HC$_3$N are shown in Fig. \ref{fig:nc}. We see that $n_c$ increases from 10$^3$ cm$^{-3}$  for $J_u$=1 to more than 10$^6$ cm$^{-3}$ for $J_u$=20. 
For $J_u>15$, we note that the critical density is higher than 10$^6$ cm$^{-3}$. This region corresponds to energies $E_u$ greater
than 100 K, which is the critical value at which the population diagram of Fig. \ref{fig:rot} exhibits a change in slope towards
higher excitation temperatures. For $n(H_2)<10^6$, this may imply that the excitation of high-$J$ levels is affected considerably by continuum radiation.\\

\subsubsection{Can the 500 K component exist?}

Infrared observations by \citet{evans03} report a dust temperature of 85 K. This value was inferred by comparing the
IR fluxes at 60 and 100 $\mu$m, and can be combined with the total IR flux to estimate the size of the emitting region.
In their paper, Evans and collaborators find an optically thick dust source of about 0.5$''$ in size, corresponding to 70 pc at a
distance of 30 Mpc. 

Assuming that our 500 K component also represents the temperature of the IR continuum, the required source size $\theta$ for the
emitting region can be estimated by the relation $L_{m IR}\propto\theta^ 2T^ 4$, which infers $\theta$=$0.01''$. This corresponds to
a linear diameter of 1.45 pc. This is an upper limit to the size of the emitting region, where it the origin of all the observed IR flux. 
Our interpretation of the molecule's excitation then implies that an extremely compact radiation source exists in the core of NGC~4418. This may correspond to a deeply embedded AGN, heating the surrounding high column of dust to
the required high temperatures. 
A hot 500 K component would peak at 10 $\mu$m, right on top
of the silicate absorption band, which in NGC~4418 is one of the deepest ever measured in a luminous IR galaxy \citep{spoon07}.
This high temperature feature might be hidden in the silicate dip and not be evident in the SED. Furthermore, \citet{lahuis07} find gas
temperatures
from the 14 $\mu$m absorption of HCN to be 300 K in the core of NGC~4418. This would require the continuum source to be hotter than 300 K gas observed in absorption.

\subsection{Models of the excitation of HC$_3$N - optically thick case}
\label{sec:models}
If we abandon the notion that HC$_3$N is optically thin, we show in Appendix~\ref{sec:opacity}
that optical depth effects may allow us to fit a single temperature (with however significant scatter)
to the v=0 transitions. This lowers the excitation temperature of the high-J levels, but leaves us with extreme
column densities of HC$_3$N and large optical depths that should result in strong lines in the $^{13}$C
isotopic variants of HC$_3$N. \\ Assuming a Galactic C/$^{13}$C  abundance ratio, these lines should have intensities comparable to the  emission of the main isotopomer. However, measurements of the 3 mm band  obtained using the new EMIR receiver at IRAM 30m (Costagliola et al., {\it in prep.}) show that the emission from H$^{13}$CCCN J=10-9 is at least 10 times fainter than that from the main $^{12}$C variant.
Even when attempting to correct for optical depth effects, we still have trouble fitting all the points
to a single temperature and the column densities required ($N$(HC$_3$N)$>10^{17}$ cm$^{-2}$)
imply HC$_3$N abundances in excess of those of HCN and HCO$^+$, starting to approach those of CO. 

\subsection{HC$_3$N abundance}

From the population diagram in Fig \ref{fig:rot}, the v$_7$ component has a column density of 8.7$\times10^{15}$ cm$^{-2}$, which can be assumed to be a lower limit to the total HC$_3$N column. \\ The hydrogen column is estimated from $^{13}$CO {J=1-0 } and {J=2-1} measurements, taken in 2007 and 2008 at IRAM 30m (Costagliola et al., {\it in prep.}). A basic LTE  interpretation of $^{13}$CO intensities leads to a source-averaged column density N($^{13}$CO)$\simeq$5.2$\times10^{16}$ cm$^{-2}$. 
Assuming a C/$^{13}$C ratio of 50, typical of starburst galaxies \citep[e.g., ][]{henkel93, mao2000} and applying a N(CO)/N(H$_2$)=$10^{-4}$ conversion \citep{blake87}, we obtain a molecular hydrogen column N(H$_2$)$\simeq2.6\times10^{22}$ cm$^{-2}$. This estimate is consistent with the value of 7.7$\times10^{22}$ cm$^{-2}$ derived by \citet{lahuis07} from silicate absorption in {\it Spitzer} spectra. The resulting HC$_3$N abundance is then X[HC$_3$N]=N(HC$_3$N)/N(H$_2$)=3.4$\times10^{-7}$.\\
In our calculations we assume that  the $^{13}$CO emission is optically thin and effectively traces the hydrogen column. The observed $^{12}$CO/$^{13}$CO J=1-0 ratio in this galaxy is about 20, implying only moderate $^{12}$CO opacities. It is therefore reasonable to assume optically thin emission for $^{13}$CO. 
% This is also supported by our observations of C$^{18}$O, obtained at IRAM 30m in July 2007, which give a lower limit for the $^{13}$CO/C$^{18}$O ratio of about three. This is a typical value for optically thin $^{13}$CO emission in starburst galaxies \citep{aalto95}. 
%Moreover, $^{13}$CO emission is very narrow, with line widths of the order of 70 km/s. This could indicate that it is not originating from the very central region of the galaxy, but could instead be more extended than our 2$''$ source. This would lead to even greater HC$_3$N abundances.}

\subsection{HC$_3$N in the Galaxy}
\label{sec:inthegalaxy}

In our Galaxy, HC$_3$N is often
associated with hot cores, i.e., dense and warm regions around newborn massive OB stars. Here, large
dust columns (A$_V\simeq$1000) provide shielding against UV radiation
from the central object. The first interstellar detection of vibrationally excited HC$_3$N was by \citet{clark76} 
toward the Orion hot core and an extensive survey of vibrationally excited lines in other
Galactic sources was performed by \citet{wyr99}. Bright HC$_3$N emission has also been found towards circumstellar envelopes
and regions around planetary nebulae \citep{pardo04}.

Our estimate of the HC$_3$N abundance has the same order of magnitude as that found by \citet{devicente00} towards hot cores in Sgr~B2, one of the most active regions of high mass star formation in the Milky Way. An enhancement of HC$_3$N abundance  in hot cores was also observed in Orion~A \citep{rodriguez98}, where the abundance is higher by about one order of magnitude than that of the diffuse warm gas, reaching values close to 10$^{-8}$. In the same complex, HC$_3$N strongly anticorrelates with HII regions, where it is likely to be destroyed by the strong UV radiation.\\
The estimated HC$_3$N abundance for NGC~4418 thus resembles the properties of Galactic hot cores, where the highest concentration of the molecule has been observed. It is quite remarkable that this appears to be a global property of the galaxy, especially considering that $^{13}$CO emission might be more extended than HC$_3$N and thus the estimated abundance might be only a lower limit to its true value.

\subsection{HC$_3$N in other galaxies}
\label{sec:othergal}

One reason for selecting  NGC~4418 to perform a deeper study of its HC$_3$N component was the unusual
luminosity of its HC$_3$N 10--9 line. Combining observational data with data in the literature, \citet{lind} classified galaxies as {\it HC$_3$N-luminous} when the HC$_3$N 10--9 line was at least
15 \% of the HCN 1--0 line intensity. He found that only a few (5) of 30 galaxies  in his sample
fitted this category. NGC~4418 belongs to this exclusive class of HC$_3$N-luminous objects,
having an HC$_3$N/HCN line intensity ratio of 0.4.
In a high resolution study of IC~342, \citet{meier05} find that HC$_3$N emission correlates well
with the 3 mm continuum emission, and avoids regions of high UV radiation. For example, they find
a spatial anticorrelation between HC$_3$N and the PDR-tracer C$_2$H. The conclusion is that HC$_3$N traces
young star-forming regions. On the scale of the main molecular complexes, the HC$_3$N/HCN ratio is 0.28, which is still
lower than the global value for NGC~4418.
\citet{wang04} report a summary of HC$_3$N abundances in the nearby starbursts NGC~253 and M~82 as
well as the Seyfert-2 NGC~4945, where these range from X=10$^{-9.0}$ to 10$^{-7.9}$.
For these three galaxies, Lindberg lists the HC$_3$N/HCN intensity ratio to be less than 0.1 for NGC~253 and NGC~4945
and for M~82 it is less than 0.2. We note that these are central and not global ratios, these three galaxies being extended compared to the single-dish beam.

\subsection{Outlook}

To confirm the presence of a 260 K component, we need carefully
calibrated submm data of the high-$J$ transitions. To more tightly constrain
the excitation and spatial extent of HC$_3$N, high resolution interferometric
observations are important. A hot, 500 K, dust component should be identifiable by
high resolution IR observations.

In a previous paper we proposed that HC$_3$N could be used as an indicator of AGN
since HC$_3$N should be destroyed by both ions and UV radiation. In a follow-up paper, we will present models
of the impact of XDRs and PDRs on the HC$_3$N abundances and discuss the survival and formation of HC$_3$N
in these environments. We do not discuss in this paper the possibility that the hot component is
not excited by a radiation source, but that the temperature instead reflects the formation and destruction
processes of HC$_3$N, thus indicating that they must be taken into account when modeling the excitation
of the molecule. 
We are aware of the fact that a non-LTE analysis would be required to  properly address the problem of the molecule's excitation. A complete analysis of the excitation and the chemistry of HC$_3$N will be  discussed in a follow-up paper.

\section{Conclusions}

We have confirmed the first  extragalactic detection of vibrational lines of HC$_3$N.
We have detected 6 different rotational transitions ranging from $J$=10--9 to $J$=30--29 in the ground vibrational state, plus a tentative detection of the $J$=38--37 line. We have also detected 7 rotational transitions of the vibrationally excited states $v_6$ and $v_7$, with angular momenta ranging from $J$=10--9 to 28--27.  In the optically
thin regime, we find that the $v=0$
transitions can be reproduced by models of two temperatures, 29 K and 265 K, while the $v_7$
lines can be fitted by a temperature of 91 K. The vibrational temperature, fitted to the J=25-24 transitions is, 519 K. By allowing the column density to vary between the different temperature components, we inferred HC$_3$N column densities of 8$\times10^{14}$ cm$^{-2}$ for the vibrational ground level, and 9$\times10^{15}$ cm$^{-2}$ for the vibrational $v_7$=1 transitions.\\
The excitation of the HC$_3$N molecule responds strongly to the intense radiation field and the presence of
 warm, dense gas and dust at the center of NGC~4418. The intense HC$_3$N line emission is a result of both high
abundances and excitation. The HC$_3$N excitation and abundances seem similar to those
found for hot cores in Sgr B2 in the Galactic Center. This
implies that the nucleus of NGC~4418 has properties in
common with Galactic hot cores. It cannot be excluded that the hot (500 K) component
may be associated with a buried AGN.

\begin{acknowledgements}

 We thank the staff at the IRAM 30m telescope for their kind help and
support during our observations. Furthermore, we would like to thank the 
IRAM PC for their generous allocation of time for this project. 

This research was supported by the EU Framework 6 Marie Curie Early Stage Training programme under contract number MEST-CT-2005-19669 "ESTRELA" and by the European Community Framework Programme 7, Advanced Radio Astronomy in Europe, grant agreement no. 227290, "RadioNet".

\end{acknowledgements}

\bibliographystyle{aa} % style aa.bst

\begin{appendix}
\section{Opacity effects}
\label{sec:opacity}

\begin{figure}
\centering
\includegraphics[width=.4\textwidth,keepaspectratio]{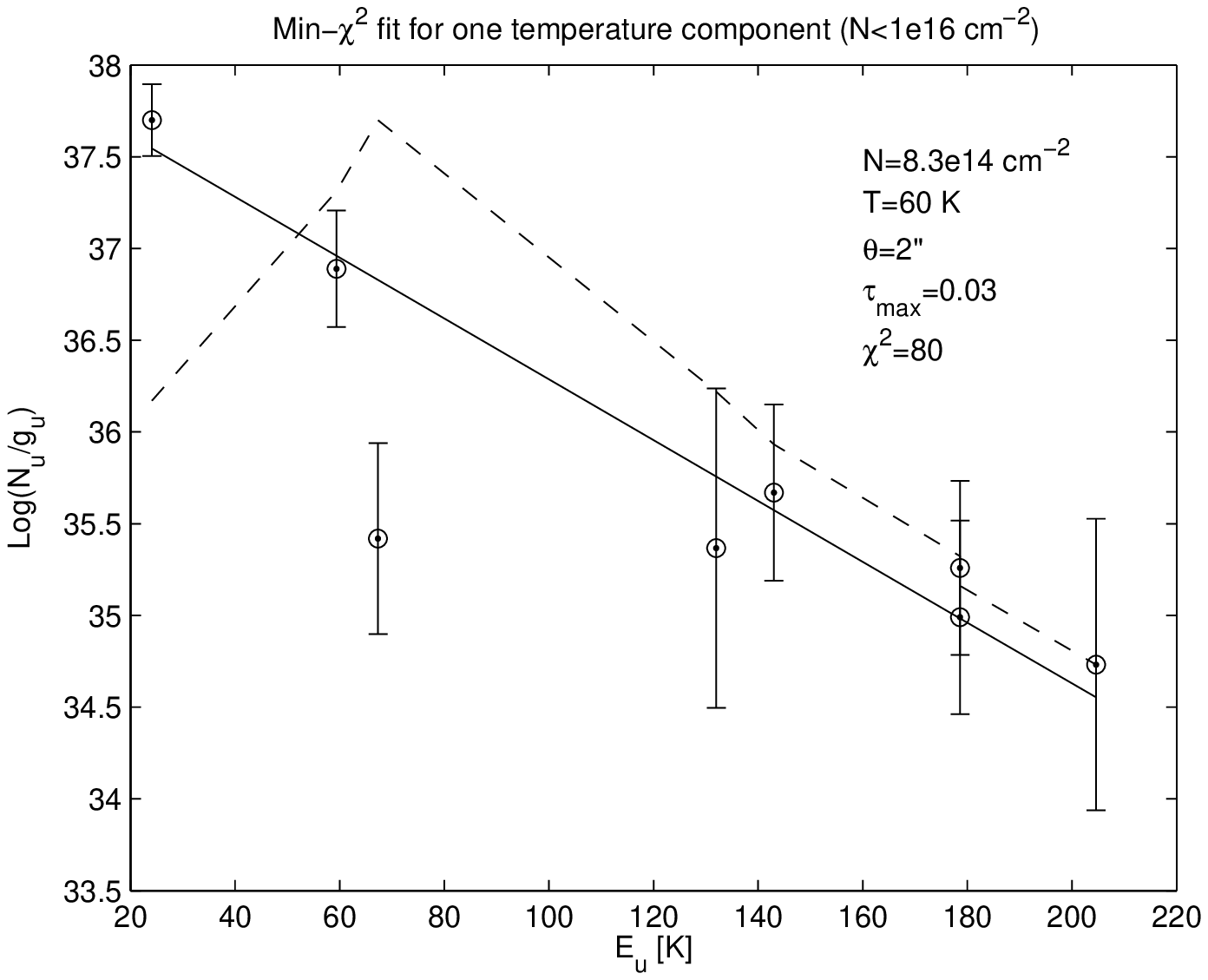}\\
\includegraphics[width=.4\textwidth,keepaspectratio]{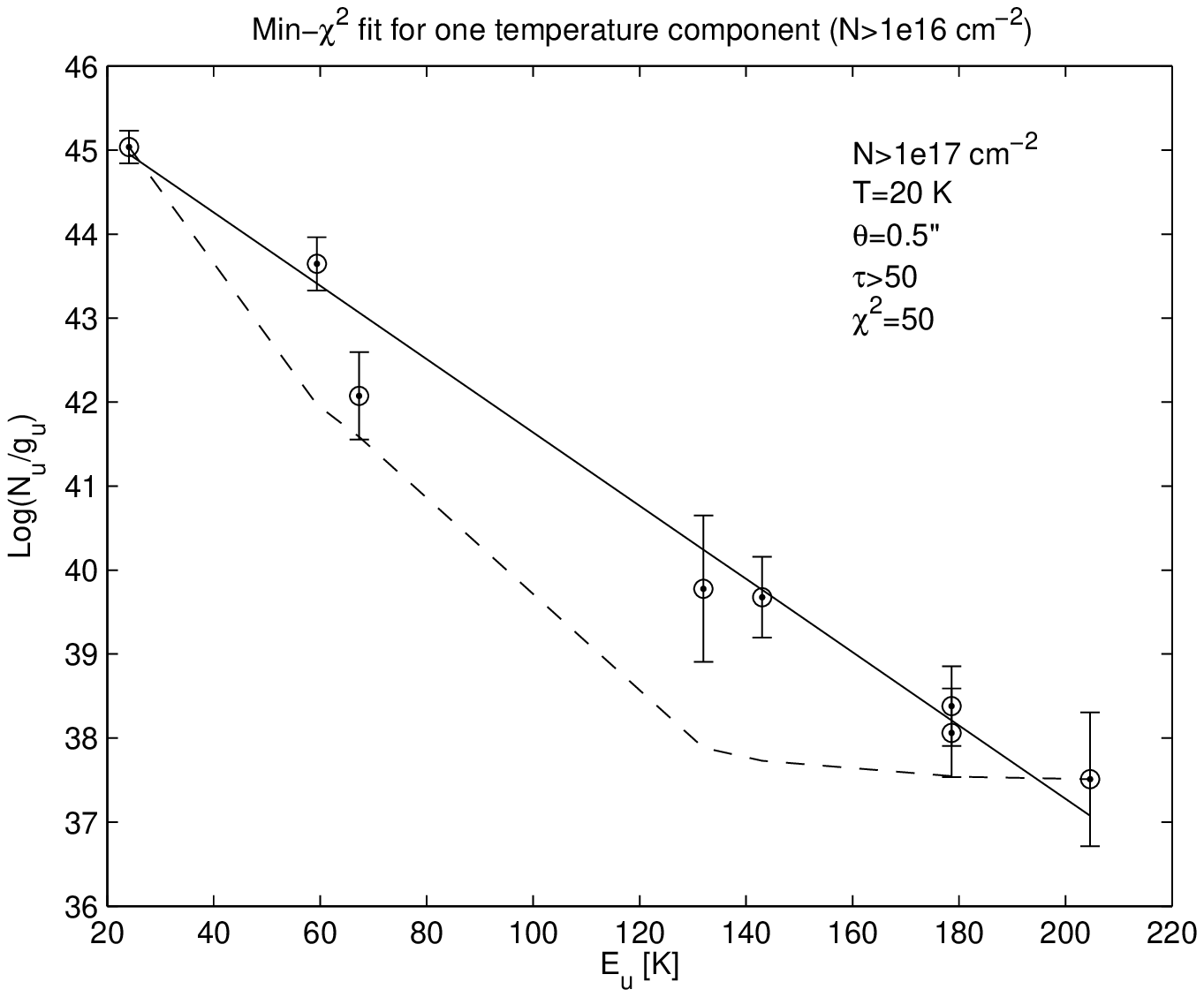}
\caption{Fit of HC$_3$N $v$=0 intensities including optical depth effects. {\it Top:}  Fit for HC$_3$N column density smaller than
10$^{16}$ cm$^{-2}$ (normalized $\chi^2$=80). {\it Bottom:} Fit for HC$_3$N column density larger than 10$^{16}$ cm$^{-2}$ (normalized $\chi^2$=50).
In both fits, the source size $\theta$ could vary between 0.1 and 2 arcseconds. The dashed line represents the optical depth for the different transitions, in arbitrary units.}
\label{fig:fitv0}
\end{figure}

The effects of a finite optical thickness of the observed lines on the population diagram have been investigated by different
authors, e.g., \citet{popdiag} and \citet{wyr99}. \\If we were to define $C_\tau=\tau/(1-e^{-\tau})$ as the correction due
to optical depth, Eq. \ref{eq:pop} would read
 \begin{equation}
%\begin{center}
\label{eq:popthick}
N_u=\frac{8\pi k \nu^2 W}{h c^3 A_{u\ell}}C_\tau\equiv N^{thin}C_\tau.
%\end{center}
\end{equation}
For $\tau<<1$, $C_\tau\simeq 1$ and $N_u\simeq N^{thin}$, but for $\tau>1$, we have $C_\tau\simeq \tau$
and the observed line intensities lead to an underestimate of the derived column density if  no correction is applied.\\
For a linear molecule, the optical depth of a $J \rightarrow J-1$ transition is given by 

\begin{equation}
%\begin{center}
\label{eq:tau}
%\tau=\frac{A_{u\ell}c^3}{ 8 \pi \nu^3}\frac{N}{Z\Delta v}e^{-\alpha J(J+1)}\left(e^{2 \alpha J}-1\right)(2J+1)
\tau=\frac{8\pi^3\mu^2}{3h}\frac{N}{Z\Delta v}Je^{-\alpha J(J+1)}\left(e^{2\alpha J}-1\right),
%\end{center}
\end{equation}
where $\mu$ is the dipole moment of the molecule and $\alpha\equiv h B_0/k T_{\rm ex}$, where $B_0$ is the rotational constant.
If $k T_{\rm ex} >> hB_0 $, the partition function can be approximated by $Z \simeq a^{-1} $ \citep{popdiag}. This is often the
case for HC$_3$N in interstellar space, since B$_0$=4.55 GHz and the previous condition becomes $T_{\rm ex}>>$0.2 K. \\
The lack of detected emission from $^{13}$C-variants of HC$_3$N (see Sect. \ref{sec:models}) prevents us from obtaining a direct estimate of the optical depth. We therefore
used Eqs. \ref{eq:popthick} and \ref{eq:tau} to fit the observed intensities via a minimum-$\chi^2$ method. We focused in particular on the $v$=0 lines, to find out whether the change in slope could be due to optical depth effects rather than a
high-temperature population.
The result of the fit is shown in Fig. \ref{fig:fitv0}. The free parameters were excitation temperature, column density, and
source size, which were varied across a wide range of values.

We obtain two possible solutions, one at N=8.3$\times 10^{14}$ cm$^{-2}$ and one at N$>10^{17}$ cm$^{-2}$. Both panels in Fig. \ref{fig:fitv0}
show that a single component does not fit the observed intensities. When we reach the optically thick regime (around 10$^{16}$-10$^{17}$ cm$^{-2}$)
the fit improves very slowly with increasing column density. However, even for extremely large columns (e.g. 10$^{25}$ cm$^{-2}$), the
normalized $\chi^2$ is still high (around 20).  Thus the change in slope of the population diagram cannot be explained by optical depth effects.\\

An estimate of the optical depth of the observed lines can be obtained by comparing their brightness temperature (T$_{b}$ in Table \ref{tab:lines}) to the excitation temperature. In general, for a spectral line we have $T_{b}=(T_{ex}-T_B)(1-e^{-\tau})$. Neglecting the contribution from the background $T_B$, the optical depth is given by $\tau=-\ell n(1-T_{b}/T_{ex})$. Considering an excitation temperature of 30 K for the first three $v$=0 lines in Table \ref{tab:lines}, we derive $\tau<0.06$, which is consistent with our optically thin scenario.

\end{appendix}

\end{document}